\documentclass[preprint2]{proto}
\usepackage{times}
\newcommand{\refs}{\par\noindent\hangindent=1pc\hangafter=1}
\begin{document}

\title{\textbf{\LARGE Atmospheres of Extrasolar Giant Planets}}

\author {\textbf{\large Mark S. Marley and Jonathan Fortney}}
\affil{\small\em NASA Ames Research Center}

\author {\textbf{\large Sara Seager}}
\affil{\small\em Carnegie Institute of Washington}

\author {\textbf{\large Travis Barman}}
\affil{\small\em University of California at Los Angeles}

\begin{abstract}
\baselineskip = 11pt
\leftskip = 0.65in 
\rightskip = 0.65in
\parindent=1pc

{\small The key to understanding an extrasolar giant planet's spectrum--and hence its detectability and evolution--lies with its atmosphere.  Now that  direct observations of thermal emission from extrasolar giant planets are in hand, atmosphere models can be used to constrain atmospheric composition, thermal structure, and ultimately the formation and evolution of detected planets.  We review the important  physical processes that influence the atmospheric structure and evolution of extrasolar giant planets and consider what has already been learned from the first generation of observations and modeling.  
We pay particular attention to the roles  of cloud structure, metallicity, and atmospheric chemistry in affecting detectable properties through {\it Spitzer Space Telescope} observations of the transiting giant planets. Our review stresses the uncertainties that ultimately limit our ability to interpret EGP observations.  Finally we will conclude with a look to the future as characterization of multiple individual planets in a single stellar system leads to the study of comparative planetary architectures.
 \\~\\~\\~}

\end{abstract}  

\section{\textbf{INTRODUCTION}}

Atmospheres of planets serve as  gatekeepers, controlling the fate of incident radiation  and regulating the loss of thermal energy.  Atmospheres are also archives, preserving gasses that reflect the formation and the evolution of a planet. Thus a complete characterization of an extrasolar giant planet entails understanding its thermal evolution through time, bulk and atmospheric composition, and atmospheric structure.  To date transit spectroscopy has probed the chemistry of the upper atmosphere of one EGP, and broad band measurements of the flux emitted by two extrasolar giant planets were reported in 2005. Many more such observations will follow as we await the direct imaging and resultant characterization of many EGPs around nearby stars.

This review focuses on the physics of giant planet atmospheres and the models which describe them.  We first approach these planets from a theoretical perspective, paying particular attention to those aspects of planetary models that directly relate to understanding detectability, characterization, and evolution. We stress the modeling uncertainties that will ultimately limit our ability to interpret  observations.   We will review the observations of the transiting giant planets and explore the constraints these observations place on their atmospheric structure, composition, and evolution.    Unlike purely radial velocity detections, direct imaging will allow characterization of the atmosphere and bulk composition of extrasolar planets, and provide data that will shed light on their formation and evolution through time.  We will explore what plausibly can be learned from the first generation of EGP observations and discuss likely degeneracies in interpretation that may plague early efforts at characterization.

\bigskip
\noindent
\section{\textbf{OVERVIEW OF GIANT PLANET ATMOSPHERES}}

The core accretion theory describing the formation of giant planets ({\em Wetherill and Steward}, 1989; {\em Lissauer}, 1993) suggests that any planet more massive than about 10 Earth masses should  have accreted a gaseous envelope from the surrounding planetary nebula.  This leads to the expectation that any massive planet will have a thick envelope of roughly nebular composition surrounding a denser core of rock and ice.  For this review we implicitly adhere to this viewpoint.   Because subsequent processes, such as bombardment by planetesimals, can lead to enhancements of the heavier elements, we don't expect the composition of the planetary atmosphere to precisely mirror that of the nebula or the parent star.  Observed enhancements of carbon in solar system giant planets (Figure 1), for example, range from a factor of about 3 at Jupiter to about 30 times solar abundance at Uranus and Neptune. 
\begin{figure*}
 \epsscale{1.0}
 \plotone{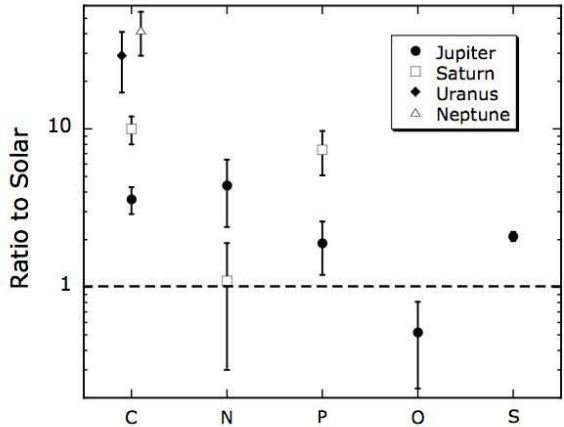}
  \caption{\small Measured atmospheric composition of solar system giant planets (neglecting the noble gasses) expressed as a ratio to solar abundance ({\em Lodders}, 2003).  Jupiter and Saturn abundances are as discussed in {\em Lodders} (2004), {\em Visscher and Fegley} (2005), and {\em Flasar et al.} (2005). Uranus and Neptune abundances are reviewed in {\em Fegley et al.} (1991) and {\em Gautier et al.} (1995).}  
 \end{figure*}

Departures from nebular abundance provide a window to the formation and evolution history of a planet. 
The near uniform enrichment of heavy elements in the atmosphere of Jupiter ({\em Owen et al.}, 1999) has been interpreted as evidence that planetesimals bombarded the atmosphere over time (e.g., {\em Atreya et al.}, 2003).  Direct collapse of Jupiter from nebular gas would not result in such a pattern of enrichment. A major goal of future observations should be to determine if most EGPs are similarly enriched in atmospheric heavy elements above that in the atmosphere of their primary stars.

Other outstanding questions relate to the thermal structure, evolution, cloud and haze properties, and photochemistry of extrasolar giant planets.    For discussion it is useful to distinguish between cooler, Jupiter-like planets and those giant planets that orbit very close to their primary stars, the `hot Jupiters'.   While the ultimate goals for characterizing both types of planets are similar, the unique atmospheres of the two classes raise different types of questions.  For the hot Jupiters, most research has focused on the horizontal and vertical distribution of incident stellar radiation in their atmospheres and the uncertain role of photochemical processes in altering their equilibrium atmospheric composition.  The available data from the transiting hot Jupiters also challenges conventional atmospheric models as their emergent flux seems to be grayer than expected.

Cooler, more Jupiter-like planets have yet to be directly detected.  Consequently most research focuses on predicting the albedos and phase curves (variation of brightness as a planet orbits its star caused by the angular dependence of atmospheric scattering) of these objects to aid their eventual detection and characterization.  As with the solar system giants, most of the scattered light reflecting from extrasolar EGPs will emerge their cloud decks.  Thus developing an understanding of which species will be condensed at which orbital distances and--critically--the vertical distribution of those condensates is required to facilitate their characterization.  Finally, for both types of planets, second order effects, including photochemistry and non-equilibrium chemical abundances can play surprisingly large roles in controlling the observed planetary spectra. 

\bigskip
\noindent
\subsection {Atmospheric Temperature and Evolution}

A key diagnostic of the thermal state of a giant planet atmosphere is the effective temperature, $T_{\rm eff}$.  The total luminosity, $L$, of a planet with radius $R$ arises from the combination of emission of absorbed incident stellar energy $\pi {\cal F}_\star$ and the intrinsic internal heat flux $L_{\rm int}$:
$$L=4\pi R^2\sigma T_{\rm eff}^4 = (1-\Lambda)\pi R^2 (\pi {\cal F}_\star) + L_{\rm int}. \eqno(1)$$
The Bond albedo, $\Lambda$, measures the fraction of incident energy scattered back to space from the atmosphere.  
The reradiation of thermalized solar photons (the first term on the right-hand side) makes up about 60\% of the total luminosity of Jupiter and Saturn.  For a hot Jupiter, 99.99\% of the planet's luminosity is due to reradiation of absorbed stellar photons (the first term on the right in the equation).  As a giant planet ages, the contribution to the total luminosity from cooling of the interior, $L_{\rm int}$,  falls as this energy is radiated away.

For a remotely detected planet, Eq. (1) can be used to constrain the planetary radius.  Given independent measurements of the total emitted infrared flux and the reflected visible flux--and assuming the internal flux is either negligible or precisely known from models--then the planetary radius can be inferred with an accuracy limited only by that of the optical and thermal IR photometry.  However for EGPs the internal flux will not be well constrained since this quantity varies with the age and mass of the planet, both of which will not be known perfectly well.  Furthermore, for solar system planets, models only predict well the internal luminosity of Jupiter.  Standard cooling models under-predict the internal luminosity of Saturn and over-predict that of Uranus by a sizeable margin.  Realistic errors in remotely sensed EGP radii will be dominated by the uncertainty in $L_{\rm int}$ and could easily exceed 25\%, particularly for objects which have high internal luminosities compared to absorbed incident radiation (e.g., planets somewhat younger or more massive than Jupiter).   Thus bulk composition inferred from the measured mass and radius will be highly uncertain. 

To aid in the interpretation of observations of a given object, modelers frequently compute a one-dimensional, globally averaged temperature profile (connecting temperature to pressure or depth vertically through the atmosphere).  Reflected and emitted spectra can be computed from such a profile (Section 2.2).  However since the fraction of the incident stellar radiation varies over a globe, one must first choose what fraction, $f$,  of the normal incidence stellar constant should strike the upper layers of a  one-dimensional atmosphere model. Setting $f=1$ results in a model  atmosphere that is only correct for the planet's substellar point.  Combined with an appropriate choice for the mean solar incidence angle, setting $f=1/2$ gives a day side average while $f=1/4$  gives a planet-wide average.  The latter is the usual choice for models of solar system atmospheres, since radiative time constants are typically long compared to rotation periods, allowing the atmosphere to come to equilibrium with the mean incident flux.  
Such an `average' profile may be less meaningful for tidally locked planets, depending on the atmospheric temperature, radiative time constant, and circulation. 

The internal energy of a giant planet ($L_{\rm int}$), a remnant of its formation, is transported through the bulk of the planet's fluid interior by efficient convection, as first discussed by {\em Hubbard} (1968). Whether a giant planet is at 0.05, 0.5, or 5 AU from its parent star, the rate at which this internal energy is lost is controlled by the planet's atmosphere.  In general,  the closer a planet is to its parent star, or the smaller its flux from the interior, the deeper the boundary between the atmospheric radiative zone and the convective deep interior will be (Figure 2).  Models indicate that the radiative/convective boundary is at $\sim 0.5\,\rm bar$ in Jupiter and can range from 10 to $\sim 1\,\rm kbar$ in hot Jupiters ({\em Guillot et al.}, 1996; {\em Barman et al.}, 2001; {\em Sudarsky et al.}, 2003). Cooling and contraction is slow for planets with deeper radiative zones because the flux carried by the atmosphere is proportional to the atmosphere's temperature gradient (see {\em Guillot and Showman}, 2002).

Connecting planetary age to total luminosity or effective temperature presents a number of challenges.  First, evolution models depend upon average planetary atmospheric conditions since the rate of cooling of the interior is governed by the mean energy loss of the entire planet.  For hot Jupiters, `mean' conditions may involve subtleties of radiative transport, dynamics, and convection.  {\em Guillot and Showman} (2002) have shown that cooling and contraction are hastened for models that include temperature inhomogeneities at deep levels, rather than a uniform atmosphere, given the same incident flux. Recently, {\em Iro et al.} (2005) have computed time dependent radiative models for HD 209458b, including energy transport due to constant zonal winds of up to $2\,\rm km/s$.  They find that at altitudes deeper than the 5 bar pressure level, as the timescale for the atmosphere to come into radiative equilibrium becomes very long, pressure-temperature profiles around the planet become uniform with longitude and time, and match a single `mean' profile computed using $f=1/4$.  This clearly suggests that model atmosphere grids computed with $f=1/4$ are most nearly correct for use as boundary conditions for evolution models.  Models that use $f=1/2$  (such as {\em Burrows et al.}, 2003; {\em Baraffe et al.}, 2003; {\em Chabrier et al.}, 2004), overestimate the effect of stellar irradiation on the evolution of giant planets, as they assume \emph{all} regions of the atmosphere receive the flux of the day side. 

\begin{figure*}
 \epsscale{1.4}
 \plotone{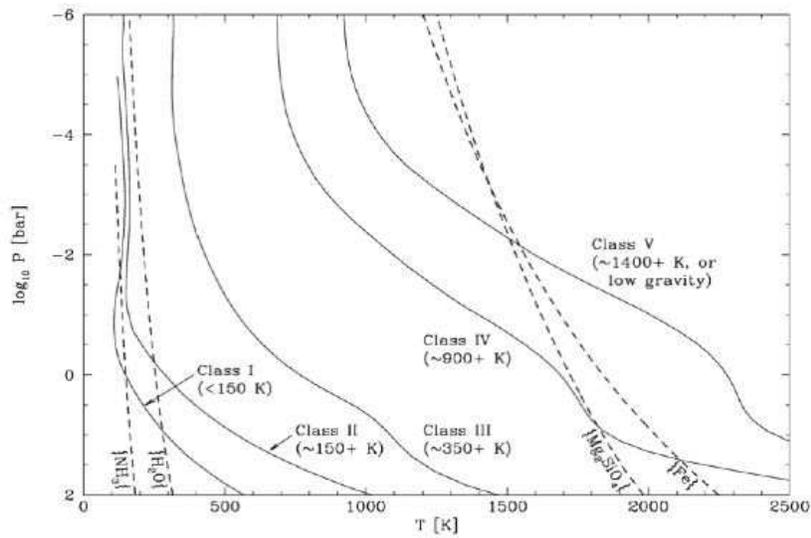}
  \caption{\small Model temperature-pressure ($T$-$P$) profiles for cloudless atmospheres illustrating the giant planet classification scheme proposed by {\em Sudarsky et al.} (2003).  Vertical dashed lines identify condensation curves for iron (Fe), forsterite ($\rm Mg_2SiO_4$), water ($\rm H_2O$), and ammonia ($\rm NH_3$) in a solar-composition atmosphere.  The base of  cloud for a given condensate is predicted to occur where the atmosphere $T$-$P$ profile crosses the species' condensation curve. The Class V planets have high iron and silicate clouds, Class III planets are relatively cloudless, and Class I planets have high ammonia clouds, like Jupiter.} 
 \end{figure*}

A second difficulty with evolution models relates to the depth at which incident stellar energy is deposited.  For Jupiter-like atmospheres (Figure 2) the deep interior of the planet is connected to the visible atmosphere by a continuous adiabat.  Thus absorbed stellar energy simply adds to the internal energy budget of the planet and--with appropriate book keeping--atmosphere models appropriate for isolated, non-irradiated objects can be used to compute the evolution.  Evolution models computed in this limit indeed work well for Jupiter ({\em Hubbard}, 1977). For hot Jupiters, however, the deep adiabat is separated by a radiative, isothermal region from that part of the atmosphere that is in equilibrium with the incident radiation (Figures 2 and 3). As discussed by {\em Guillot and Showman} (2002), using atmosphere models suitable for isolated objects as the boundary conditions for hot Jupiter evolution calculations (e.g., {\em Burrows et al.}, 2000) significantly overestimates the temperature of the atmosphere at pressures of $\sim 1\,\rm bar$, leading one to overestimate the effect of irradiation, and predict contraction that is too slow.  In contrast, {\em Bodenheimer et al.} (2001, 2003) and {\em Laughlin et al.} (2005) have computed hot Jupiter evolution models where contraction is likely too fast.  For their atmospheric boundary calculation, the atmospheric pressure at which the Rosseland mean optical depth reaches 2/3 ($\sim 1\,\rm mbar$ in their models) is assigned the planetary effective temperature, which itself is calculated after assuming a Bond albedo.  This method assumes a very inefficient penetration of stellar flux into the planet's atmosphere, compared to detailed atmosphere models.  Temperatures at higher atmospheric pressures are underestimated, leading to an underestimation of the effect of irradation.  The ideal solution is to compute individualized atmosphere models to use as boundary conditions for many timesteps in the evolutionary calculation.

\subsection{Spectra of Giant Planets}

The reflected and emitted spectra of giant planet atmospheres are controlled by Rayleigh and Mie scattering from atmospheric gases, aerosols, and cloud particles, and by absorption and emission of gaseous absorbers.  Scattering of incident light usually dominates  in the blue, giving way to absorption by the major molecular components at wavelengths greater than about $0.6\,\rm \mu m$.  The major absorbers in the optical are methane and, for warmer planets, water.  Na and K are major optical absorbers in the atmospheres of the hot Jupiters. Generally speaking, in strong molecular bands photons are absorbed before they can scatter back to space. In the continua between bands photons scatter before they are absorbed.  The continuum flux  from a given object is thus controlled by Mie scattering from its clouds and hazes and Rayleigh scattering from the column of clear gas above the clouds.  Longward of about 3 to $5\,\rm \mu$m for the cooler planets and at shorter wavelengths for the warmest, scattering gives way to thermal emission.   

The likelihood of absorption, and hence the depth of a given band, depends upon the molecular opacity at a given wavelength and the column abundance of the principal gaseous absorbers (e.g., methane and water) above a scattering layer.  The column abundance of an absorber, in turn, depends upon the gravity (known for solar system planets), the height of a cloud layer, and its mixing ratio.  Thus the spectra, even at low resolution, of EGPs are sensitive to their atmospheric temperature, metallicity, cloud structure, and mass. In principle by comparing observed spectra to models one can infer these properties from the data.   Experience with the giant planets of our solar system, however, has shown that degeneracies in cloud properties and molecular abundances can be difficult to disentangle without broad, high-resolution spectral coverage.

 As with the optical and near-infrared wavelengths, the thermal emission of EGPs is sculpted by molecular opacities.  In regions 
 of low absorption, planets brightly emit from deep, warm layers of the atmosphere. Within strongly absorbing regions flux arises from
 higher, cooler layers (unless there is a stratospheric temperature inversion).  Bright emission in the window 
 around $5\,\rm \mu$m was flagged by {\em Marley et al.} (1996) as a diagnostic of the entire class of EGPs (but see the caveat in
  Sect.~3.4 below).  This  opacity window is responsible for the well-known `five-micron hot spots' of Jupiter ({\em Westphal et al.}, 1974).  Solar system giant planets also exhibit true emission 
features arising from a temperature inversion above the tropopause, notably in the $7.8\,\rm \mu$m methane band that plays an important role in the 
stratospheric energy budget. Photochemically produced ethane and acetylene also exhibit emission in some giant planet atmospheres.

The {\it Galileo} atmosphere entry probe provided a test of the ability of remote observers to accurately measure the abundance of gases in a giant planet atmosphere.  Prior to {\it Galileo}'s arrival at Jupiter the methane abundance was estimated to lie in the range of 2.0 to 3.6 times solar at Jupiter and 2 to 6 times solar at Saturn.  {\it Galileo} measured Jupiter's methane abundance to be $2.9 \pm 0.5$ times solar (see review by {\em Young 2003}) and recent observations by {\it Cassini} have pinned Saturn's methane abundance at $10 \pm 2$ times solar ({\em Flasar et al.}, 2005; {\em Lodders}, 2004).  In both cases (at least some) remotely sensed value were accurate (e.g., {\em Buriez and de Bergh}, 1981).  Remotely determining the abundance of condensed gases, such as ammonia or water,  is more problematic and pre-{\it Galileo} measurements were not as accurate. Fortunately, ammonia will not condense in planets just slightly warmer than Jupiter. In young or more massive planets, water will be in the vapor phase as well, which should allow for more accurate abundance retrieval.

\bigskip
\section{\textbf {MODEL ATMOSPHERES}}

Model atmospheres predict the appearance of EGPs. They thus
facilitate the design of optimal detection strategies and play a role in interpreting observations.  A typical model recipe begins with assumptions about the atmospheric
 elemental composition, the atmospheric chemistry, the internal heat flow of the planet, the incident stellar flux, and various radiative transfer assumptions (e.g., is the atmosphere in local thermodynamic equilibrium?).  A cloud model for the treatment of 
atmospheric condensates and  the relevant gaseous opacities are also required ingredients.  When combined with a suitable method
for handling atmospheric radiative and convective energy transport,  the modeling process yields the thermal structure of the 
atmosphere and the reflected and emitted spectrum.  Of course as in any recipe, models of giant planet atmospheres are only as good as the quality of the ingredients and the assumptions.  Neglected physical processes, including some that might initially seem to be of only second order importance can in fact have first order effects and spoil the predictions, at least in certain spectral regions.  In this section we summarize typical inputs into atmosphere models and discuss their relative contributions to
 the accuracy of the final product.  Figure 3 provides a comparison of hot Jupiter profiles computed by three different groups.  The differences between the profiles give an indication of the uncertainty in our understanding of these atmospheres just due to varying modeling techniques.

\begin{figure*}
 \epsscale{1.5}
 \plotone{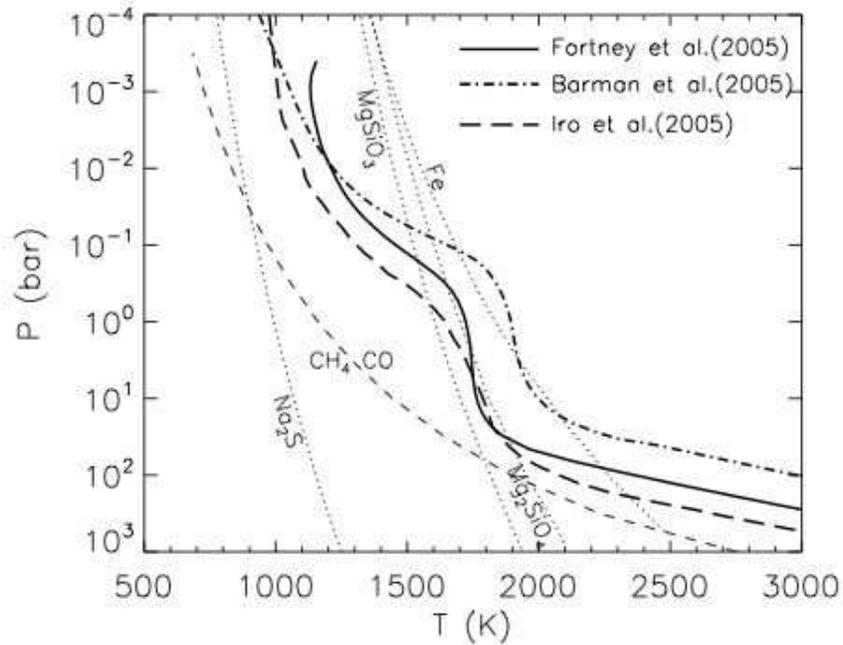}
  \caption{\small Comparison of model atmosphere profiles for hot Jupiter HD209458 computed by three groups (labeled).  Each profile assumes somewhat different intrinsic luminosity, $L_{\rm int}$, so differences at depth are not significant.  All three profiles assume global redistribution of incident energy, or $f=4$. Differences at lower pressure presumably arise from different methods for computing chemical equilibria, opacities, and radiative transfer assumptions.  The spread in models provides an estimate of the current uncertainty in modeling these objects. In addition to the internal luminosity the size of the deep isothermal layer depends upon the behaviors at high pressure of the opacities of the major atmospheric constituents, which are poorly known.  }  
 \end{figure*}

\bigskip
\noindent
\subsection{Chemistry}

Perhaps the most elemental input to an atmosphere model is the assumed composition of the atmosphere.  To date most models of EGP atmospheres have assumed solar composition.  However the best estimate of ``solar'' has changed over time (see {\em Lodders}, 2003 for a review) and of course the composition of the primary star will vary between each planetary system.  Notably the carbon and oxygen abundances of the solar atmosphere remain somewhat uncertain ({\em Asplund}, 2005) and vary widely between stars ({\em Allende-Prieto et al.}, 2002).

 Given a set of elemental abundances, a chemical equilibrium calculation provides the abundances of individual species at any given temperature and pressure.  Several subtleties enter such a calculation, particularly the treatment of condensates.  In an atmosphere subject to a gravitational field, condensates are removed by settling from the atmosphere above the condensation level.  Thus `equilibrium' reactions that might take place were the gas to be kept in a sealed container will not proceed in a realistic atmosphere.  A canonical example (e.g., {\em Fegley and Lodders}, 1994) is that under pure equilibrium, sulfur-bearing gasses would not appear in Jupiter's atmospheres since S reacts with Fe grains at low temperature to form FeS.  In fact, when iron condenses near 1600 K in Jupiter's deep atmosphere, the grains fall out of the atmosphere allowing S to remain as $\rm H_2S$ at low temperature.  The removal of condensates from the atmosphere by sedimentation is sometimes termed `rainout,' but this term can be confusing since rigorously `rain' refers only to the sedimentation of 
liquid water.  Some early brown dwarf models did not properly account for sedimentation, but most recent modeling efforts do include this effect  (see {\em 
Marley et al.}, 2002 for a more complete discussion).   For a recent, detailed review of the atmospheric chemistry of EGPs and brown dwarfs see {\em Lodders 
and Fegley} (2006).

Finally photochemistry, discussed further below, can alter atmospheric composition.  Trace gasses produced by the photolysis of methane in Jupiter's atmosphere, for example, are both important UV absorbers and emitters in the thermal infrared.  As such they play important roles in the stratospheric energy balance.  Photochemical products may include important absorbers or hazes that may substantially alter the the spectra of EGPs and cloud their interpretation.  {\em Yelle} (2004) thoroughly discusses the photochemistry of the upper atmosphere of EGPs and predicts thermospheres heated to over 10,000 K by the extreme ultraviolet flux impinging on the top of these planets' atmospheres. The high temperatures drive vigorous atmospheric escape by hydrogen, producing an extended cloud surrounding the planet that has been observed in transit by {\em Vidal-Madjar et al.} (2003) for HD 209458 b.  Despite the high escape flux, a negligible fraction of the total mass of the planet escapes over time ({\em Yelle}, 2004).

\bigskip
\noindent
\subsection{Opacities}

For the temperature-pressure regimes found in the atmospheres of all but the hottest extrasolar giant planets, the most important gaseous absorbers are $\rm H_2O$, $\rm CH_4$, $\rm NH_3$, Na, K.  In addition the pressure-induced continuum opacity arising from collisions of $\rm H_2$ with $\rm H_2$ and He is 
particularly important in the thermal infrared.  Other species found in brown dwarf atmospheres play a role in the hottest planets orbiting close to their primary stars.  {\em Freedman and Lodders} (2006) review the current state of the various opacity databases used in atmospheric modeling.  For the cool atmospheres most likely to be directly imaged, the opacities are fairly well known.  The greatest shortcomings of the current opacities are the lack of  a hot molecular line list for $\rm CH_4$ and the highly uncertain treatment of the far wings of collisionally-broadened lines.  Neither is a major limitation for most  EGP modeling applications.

\bigskip
\noindent
\subsection{Clouds and Hazes}

Clouds and hazes play a crucial role in controlling giant planet spectra.  In the absence of such scattering layers, red photons would penetrate to deep layers of an EGP atmosphere 
and generally be absorbed before they could be scattered ({\em Marley et al.}, 1999), leading to very low reflectivity in the red and near-infrared.  Planets 
with bright high water clouds, for example, tend to exhibit a bright continuum from scattered starlight punctuated by a few absorption bands.  Likewise silicate and iron clouds in the atmospheres of the close-in planets play major roles in controlling their spectra ({\em Seager and Sasselov}, 1998).  Furthermore for a given cloudy planet, reflected and emitted spectra are sensitive to the vertical distribution, fractional global coverage, size distribution, 
and column number density of cloud particles. 

Unfortunately clouds are notoriously difficult to model, even in Earth's atmosphere where the representation of clouds is a leading source of uncertainty in terrestrial global atmospheric circulation models.  Real clouds are a product of upward, downward, and horizontal transport of condensible vapor and solid or liquid condensate.  Their detailed structure depends on a number of highly local factors including the availability of condensation nuclei and the 
degree of supersaturation as well as a host of microphysical properties of the condensate. Approaches applying a 1-dimensional atmosphere model to what is intrinsically a 3-dimensional problem are certainly overly simplistic.  Nevertheless, given the paucity of information, simple 1-D models currently provide the most workable approach.  
 
A number of cloud models have been developed for solar system studies.  Perhaps the most widely used has been an approach focusing on microphysical time constants developed by {\em Rossow} (1978).  An important shortcoming of such an approach is that the time constants sensitively depend upon a variety of highly uncertain factors, particularly the degree of supersaturation.  {\em Ackerman and Marley} (2001) and {\em Marley et al.} (2003) review the physics employed by the most popular cloud models. {\em Ackerman and Marley} (2001) proposed a simple 1-dimensional cloud model that accounts for vertical transport of condensate and condensible gas including a variable describing the efficiency of particle sedimentation.  This model has had success fitting the
 cloudy atmospheres of L-type brown dwarfs, but is not able to predict such quantities as fractional global cloudiness or account for the rapidity of the L- to T-type brown dwarf transition.  Other modeling approaches are discussed by {\em Tsuji} (2005), {\em Helling et al.} (2004), and {\em Cooper et al.} (2003) which range from purely phenomenological efforts to detailed numerical microphysical models of dust nucleation.  Given the strong influence of clouds on EGP spectra, {\em Sudarsky et al.} (2000, 2003) have suggested 
classifying EGPs on the basis of which cloud layers are present or absent from the visible atmosphere (see Sect.~3.7).  The suggestion is appealing but might be difficult to apply
 in practice for transitional cases, hazey planets, or for objects with only limited spectral data.

\begin{figure*}
 \epsscale{1.5}
 \plotone{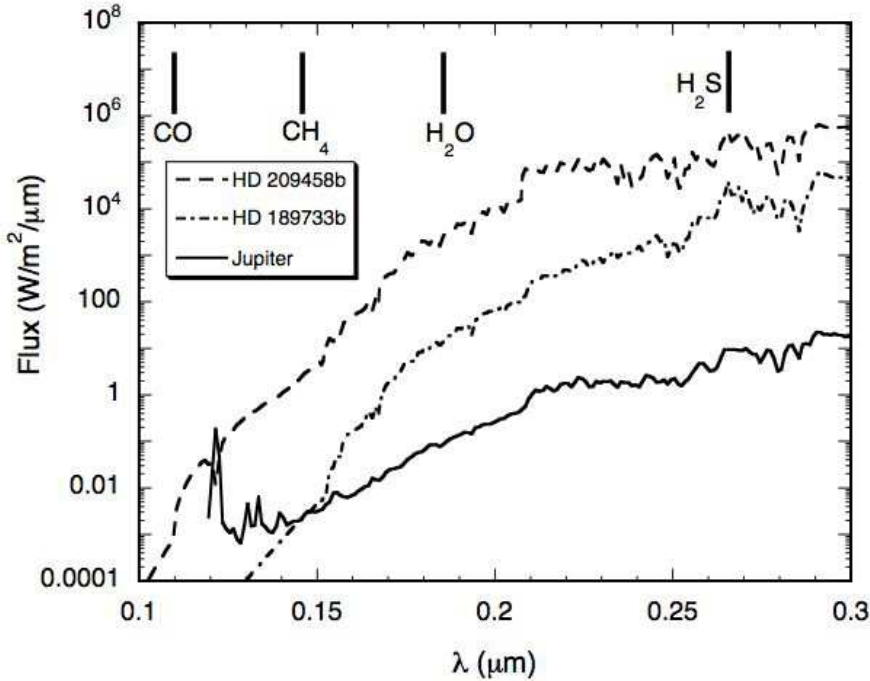}
  \caption{\small Incident flux at the top of the atmospheres of several transiting planets compared to that received by Jupiter.  Vertical lines denote the approximate maximum wavelengths at which various molecules can be dissociated.  Incident spectra at HD 149026 b and TrES-1 are similar to HD 209458 b and HD 189733 b (G4.5V), respectively, and are not shown for clarity.  Model stellar spectra from Kurucz (1993).}
   \end{figure*}

   Perhaps an even more challenging problem is atmospheric photochemical hazes.  All of the solar system giant planets
 are strongly influenced by hazes produced by the ultraviolet photolysis of methane.  Figure 4 compares the incident stellar fluxes at two transiting planets with that received by Jupiter.  The maximum wavelengths at which ultraviolet photons can photolyze various molecules are shown.  At Jupiter, solar Lyman-$\alpha$ is an important contribution of the far UV flux.  Although the primary stars of the hot Jupiters may lack substantial Lyman-$\alpha$ flux, given the proximity of the planets the integrated continuum radiation capable of photolyzing major molecules is comparable to or greater than that received by Jupiter. 
At Jupiter methane photolysis is the main driver of photochemistry since water and $\rm H_2S$ are trapped in clouds far below the upper atmosphere.  In the atmospheres of hot Jupiters, this will not be the case and these molecules will be rapidly photolyzed, perhaps providing important sources for photochemical haze production. Ultimately haze optical depths depend upon production rates, condensation temperatures, microphysical processes, and 
 mixing rates in the nominally stable stratosphere that in turn depend upon the atmospheric structure and poorly understood dynamical processes.  To date only {\em Liang et al.} (2004) have considered this issue and then only in the context of hot Jupiter atmospheres, which they found to be too warm for condensation of photochemical hydrocarbon compounds. They did not consider O- or S-derived photochemical products.  Since optically thick hazes can substantially alter the idealized EGP spectra and phase functions computed to date, much more work on their production is needed.  In any case, disentangling the effects of clouds, hazes, and uncertain atmospheric abundances in the atmospheres of EGPs will likely require high quality  spectra obtained over a large spectral range.

\bigskip
\noindent
\subsection{Dynamics and Mixing}

An important limitation to conventional 1-dimensional models of mean atmospheric structure is the neglect of vertical mixing.  Vertical transport 
plays an important role when the dynamical time scale is short compared to a particular chemical equilibrium timescale, as is the case for CO in the atmosphere of Jupiter
 ({\em Prinn and Barshay}, 1977; {\em Fegley and Prinn}, 1988; {\em Yung et al.}, 1988) and cool T-dwarfs ({\em Fegley and Lodders}, 1996; {\em Griffith and Yelle}, 1999; {\em Saumon et al.}, 2003; {\em Golimowski et  al.}, 2004).  While methane is most abundant in the visible atmospheres, in the deep atmosphere, where temperatures are higher, 
 the abundance of CO is 
substantially larger.  Since the C-O bond is very strong, the conversion time from CO to $\rm CH_4$  in a parcel of rising gas is correspondingly long. 
 This allows vertical mixing through the atmosphere to transport CO from the deep atmosphere to the visible atmosphere.  CO absorbs strongly in the $M$ photometric band and 
excess CO in T dwarf atmospheres depresses the flux in this window region by up to a magnitude below that predicted by pure 
chemical equilibrium models ({\em Saumon et al.}, 2003).  If this mechanism also depresses the flux of cool extrasolar giant planets the utility of this spectral region for planet detection 
may not be as great as predicted (e.g., {\em Burrows et al.}, 2001).

Another interesting effect is that of atmospheric dynamics.  At lower pressures the radiative timescales are shorter than at higher pressures.  For a tidally locked hot Jupiter, this will likely mean that that the upper atmosphere quickly adjusts to the flux it receives from the parent star, but deeper layers ($P>1\,\rm bar$) will adjust much more sluggishly, and the dynamic transport of energy will be important.  This is only beginning to be studied in detail ({\em Showman and Guillot}, 2002; {\em Cho et al.}, 2003; {\em Burkert et al.}, 2005; {\em Cooper and Showman}, 2005).  Infrared observations as a function of planetary phase, with the {\it Spitzer Space Telescope} and perhaps other platforms, will enable constraints to be placed on atmospheric dynamics of HD 209458b and other planets.

\bigskip
\noindent
\subsection{Albedos and Phase Curves}

Albedos are often of interest as they allow for a simple parameterization of the expected brightness of a planet. Spectra of outer solar system planets with atmospheres are commonly reported as Ôgeometric albedo' spectra, which is simply the reflectivity of a planet measured at opposition.  Other albedo definitions include the wavelength averaged geometric albedo and the Bond albedo, $\Lambda$ in Eq. (1), which measures the ratio of scattered to incident light.  Unfortunately the extrasolar planet literature on albedos has become somewhat muddled and terms are not always carefully defined. Generic ``albedos'' are often cited with no definition.  Yet different albedo varieties can differ from each other by several tenths or more.  For example the commonly referenced ÔLambert sphereÕ has a geometric albedo of 2/3 while an infinitely deep Rayleigh scattering atmosphere would have a wavelength-independent geometric albedo of 3/4, yet both have $\Lambda = 1$. The two differ in the angular dependence of their scattered radiation. For absorbing atmospheres the Bond albedo depends upon the incident spectrum. Since proportionately more red photons are absorbed than blue photons (which tend to scatter before absorption), an identical planet will have a different Bond albedo under the light of a red star rather than a blue one even though the geometric albedos are identical ({\em Marley et al.}, 1999). 

	In any case more information is needed to fully predict or interpret the flux observed by a distant observer of 
an extrasolar planet.  Geometry dictates that extrasolar planets are most detectable near quadrature and not detectable at true opposition since they would be hidden by their star, thus a general description of the phase dependence of the scattered and emitted radiation is required.  Phase information on solar system giant planets has long been used to constrain cloud particle sizes and atmospheric structure.  For example {\it Voyager} 1, which did not visit Uranus but instead imaged it from afar, observed the planet at high phase angles not reachable from Earth to help constrain the scattering phase function
 of its clouds and Bond albedo ({\em Pollack et al.}, 1986). {\em Dyudina et al.} (2005) recently relied upon {\it Voyager} observations to derive phase curves of 
Jupiter and Saturn.  {\em Marley et al.} (1999) computed phase curves for model EGPs by relying upon scattering tables computed by {\em Dlugach and Yanovitskij} (1974). More recently {\em Sudarsky et al.} (2005) presented a suite of model phase curves for EGPs.  The differences between their model calculations and the observed phase curve Jupiter (compare their figures 4 and 6) demonstrates that interpretation of specific planets will always
 be challenging since the specifics of particle size and composition, hazes, and overall atmospheric structure will likely make each giant planet discovered
 unique.

\bigskip
\noindent
\subsection{Atmosphere Models}

Although pioneered by {\em Kuiper} (1952), giant planetary atmosphere modeling entered the modern era with the work of {\em Trafton} (1967) and {\em Hogan et al.} (1969).  Following the {\it Voyager 1} and {\it 2} traverses of the outer solar system substantially more complex models were developed to explore the atmospheric energy budgets, the thermal structure, and the reflected and emitted spectra ({\em Appleby and Hogan}, 1984; {\em Appleby}, 1986; {\em Marley and McKay}, 1999) of each giant planet.  (Note that these are {\it forward} models that combine first principle information about planetary atmospheres to reproduce the observed atmospheric thermal structure.  There is also a very rich literature of inverse models that aid in the interpretation of specific data sets).  These authors modeled the equilibrium one-dimensional radiative-convective thermal profiles of these atmospheres, including deposition of incident radiation, by modeling the atmospheric radiative transfer given observed atmospheric abundances and cloud properties. The models generally well reproduced observed spectra and thermal structure.  This success provides an important reality check that 1-D modeling of giant planet atmospheres, given appropriate input, satisfactorily reproduces observed properties of giant planets.  Modeling extrasolar planets, however, will be more challenging: only the incident radiation is known with certainty. Atmospheric composition, cloud properties, and thermal structure and perhaps mass and radius will all have to be inferred from comparison of models to data.

{\em Burrows et al.} (2000) reviewed the scant early work on atmospheric modeling of the cooler irradiated EGPs.  Most pre-1995 investigations 
focused on studying the evolution of isolated objects or assumed gray atmospheres to estimate EGP detectability.  {\em Marley} (1998) computed exploratory spectra of irradiated giant planets and found that the presence or absence of water clouds is an important spectral and albedo marker in EGP atmospheres. Particularly in the red and infrared the presence or absence of scattering clouds can change the scattered flux by a factor of two or more, with cloudless planets being darker.

Atmosphere models specifically of the hot Jupiters were first developed by {\em Seager and Sasselov} (1998).   Subsequent work focusing on either specific objects or the class in general includes that by {\em Goukenleuque et al.} (2000), {\em Seager et al.} (2000, 2005), {\em Sudarsky et al.} (2000, 2003), {\em Barman et al.} (2001), {\em Iro et al.} (2005), {\em Burrows et al.} (2005), and {\em Fortney et al.} (2005, 2006).  As with the cooler planets, the main conclusion of this body of work is that the spectra of the hot Jupiters depends sensitively on the vertical distribution and properties of condensates.  Models that either postulate or predict high altitude iron and silicate cloud decks tend to be warmer and more Planckian in thermal emission than models with deeper cloud decks.    Hot Jupiter models and observations are considered in detail in Sections 3.4 and 4.

The most systematic surveys of model EGP spectra include the work of {\em Sudarsky et al.} (2000, 2003, 2005) and  {\em Barman et al.} (2001, 2005)  who have studied model planets of a variety of masses, ages, and orbital radii (Figure 2).  {\em Burrows} (2005) reviews and recasts much of the former work with an eye towards detectability of EGPs.  The universal conclusion of this body of work remains that molecular absorption bands and atmospheric condensates are the key diagnostics of giant planet effective temperature since giant planets cool as they age.  For those planets distant enough from their stars that atmospheric temperature is primarily controlled by the loss of internal energy, not incident flux, the progression to lower atmospheric temperature with age results in a diagnostic sequence of spectroscopic changes discussed in the next section.

Planets more massive than $5\,\rm M_J$ may be as warm as 2000 K shortly after formation, with temperatures falling well below 1000 K by a few hundred million years.  By a few billion years all planet mass objects ($< 13\,\rm M_{Jupiter}$ ({\em Burrows et al.}, 1997)) are cooler than 500 K.  
The important chemical equilibrium and condensation boundaries are shown in Figures 2 and 3.  As the atmosphere cools chemical equilibrium begins to favor first $\rm CH_4$ over CO and then $\rm NH_3$ over $\rm N_2$.  Water is present throughout this temperature range, but the molecular bands become deeper with falling temperature. 

The early part of this sequence has already been well sampled by observed L and T type brown dwarfs, the coolest of which is about 700 K.    

\bigskip
\noindent
\subsection{Spectral Signatures of EGPs}

No one discussion or  figure can hope to capture the range of temperature, metallicities, and cloud structures that likely define the entire suite of possible giant planets.  Nevertheless Figures 5 and 6 help illustrate the important physical processes that control EGP spectra and give an indication of the major spectral signatures expected in atmospheres with roughly solar composition.  These spectra are purposefully presented at moderate spectral resolution that will likely typify early direct detection spectra.

 \begin{figure*}
 \epsscale{1.5}
 \plotone{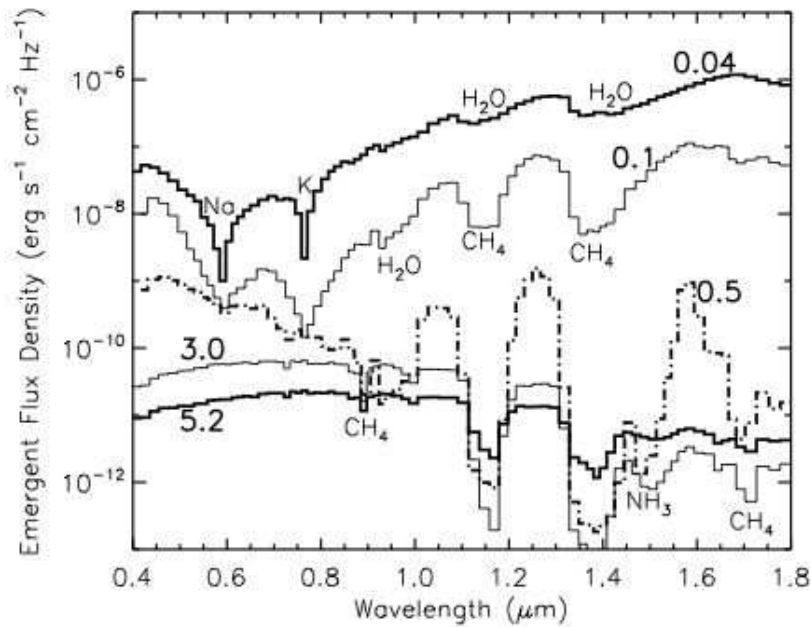}
  \caption{\small Model atmosphere spectra (computed by authors JF and MM) for giant planets roughly corresponding to the atmosphere classes shown in Figure 2.  Numbers give orbital distance $a$ from a solar type star in AU.  The top model is for a hot Jupiter ($a=0.04$ AU; $T_{\rm eff}=1440\,\rm K$; Class V).   The atmosphere is very hot with high refractory clouds. In the second model the atmosphere is cooler ($a=0.1$ AU; $T_{\rm eff}=870\,\rm K$; Class IV), the clouds are deeper, and the absorption bands are correspondingly more prominent.  At $a= 3$ AU ($T_{\rm eff}=375\,\rm K$; Class III) the atmosphere is cooler and relatively cloud free.  Remaining two curves illustrate atmosphere with water clouds ($a=0.5$ AU; Class II) and ammonia clouds ($a=5$ AU; Class I).   }  
 \end{figure*}

Figure 2 presents a set of five cloudless temperature-pressure ($T$-$P$) profiles.  Also shown are the condensation curves for iron, silicate, water, and ammonia.  The condensates expected in a given atmosphere depend upon the particular atmospheric temperature structure.  {\em Sudarsky et al.} (2003) used this atmospheric characteristic to suggest that planets be categorized by which clouds form in their atmosphere.  While this proposal has some drawbacks, discussed below, it does nicely frame the discussion of EGP atmospheres and spectra.  

The hottest EGPs orbiting most closely to their parent stars are expected to exhibit iron and silicate clouds high in their atmosphere since the atmospheric profile crosses these condensation curves at low pressures.  {\em Sudarsky et al.} term such planets Class V.  As seen in T-type brown dwarfs, Na and K are expected to be present in gaseous form and dominate the optical spectra, with water bands appearing in the near-infrared (Figure 4).  Thermal emission is an important contributor to the near-infrared flux, particularly between the strong water bands.   Cloud scattering, however, limits the band depths.

In somewhat cooler atmospheres (Class IV) the clouds form at higher pressures in the atmosphere which results in deeper absorption band depths.  In addition carbon is now found as $\rm CH_4$ rather than CO; thus methane features begin to appear in the near infrared.   

Somewhat cooler still, the effective temperature of a planet orbiting a G star at 0.5 AU would be about 375 K.  Absorption of stellar radiation keeps the atmosphere warm enough that water clouds would not form, yet the iron and silicate clouds lie far below the visible atmosphere.  Although low abundance species like $\rm Na_2S$ could form low optical depth hazes, these atmospheres (Class III) will be relatively clear with a steep blue spectral slope and deep molecular bands.  Like Class IV and V thermal emission is important beyond about $1\,\rm \mu m$.

For somewhat more distant planets, water and then ammonia condense resulting in Class II and Class I atmospheres, respectively.  Continuum flux levels are controlled by the bright cloud decks and the `giant planet bands' of methane are apparent throughout the optical, particularly the strong band at $0.889\,\rm \mu m$.  An ammonia absorption feature is detectable at $1.5\,\rm \mu m$ in Class II atmospheres, but disappears in the colder Class I since the ammonia has condensed into clouds.

Figure 5 illustrates how sensitive such predictions are to atmospheric metallicity.  Recall (Figure 1) that Jupiter's atmosphere is enhanced by a factor of 3 in carbon and Uranus and Neptune by a factor of 30.  The optical and near-infrared methane bands are highly sensitive to the methane abundance.  Continuum levels, however, vary much less since they are controlled by the cloud structure which is not as sensitive to abundance variations (the clouds are already optically thick).  A cloud-free atmosphere, however, is much darker, again illustrating the importance of clouds.

 \begin{figure*}
 \epsscale{1.5}

  \plotone{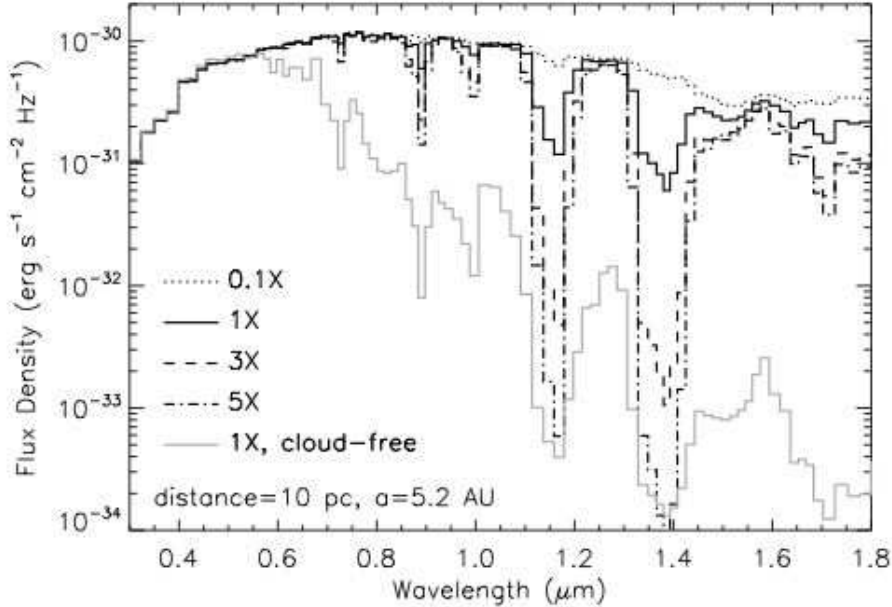}
  \caption{\small Sensitivity of model Jupiter-like spectra (computed by authors JF and MM) to changes in model assumptions. Labeled curves illustrate reflected fluxes from $1\,\rm M_J$ planets assuming, from top to bottom, 0.1 times solar abundance of heavy elements, solar abundance, 3 times solar, 5 times solar, and 1 times solar with no cloud opacity.  Note that while the continuum levels are generally unaffected by composition changes, the depths of the methane bands are highly sensitive.  The cloud-free model is much darker in the red and infrared since incident photons are far more likely to be absorbed at these wavelengths than Rayleigh scattered. }  
 \end{figure*}

\subsection{Atmospheres of the Hot Jupiters}

Despite their nickname, ``Hot Jupiters'' likely bear little resemblance to a hotter version of our own Jupiter.
Furthermore, given their extremely small orbital separations,
these planets have undoubtedly experienced a very different upbringing than their frosty jovian cousins.

Early exploratory studies into the nature of hot-Jupiters revealed that stellar
heating leads to much shallower T-P profiles than present in isolated brown
dwarfs ({\em Seager and Sasselov}, 1998; {\em Goukenleuque et al.}, 2000) (Figure 3).  Such reductions of
the temperature gradient dramatically weaken the strength of otherwise
prominent molecular absorption bands (e.g., due to water).  Also, temperatures
are high enough  that the dominant carbon based molecule is CO, unlike cooler
giant planet atmospheres which have high concentrations of CH$_4$.
Hot-Jupiter models also indicate that, even though significant amounts of
reflected optical light will be present due to Rayleigh and/or Mie scattering,
Bond albedos may be well below 0.1 ({\em Sudarsky et al.}, 2000, 2003; {\em Barman et al.}, 2001).

Even though hot-Jupiters are hot, their atmospheres are still cool enough
that molecules, liquids and even solids may form and many of the issues
mentioned above are still relevant. As always the expected atmospheric condensates
depend on the detailed thermal strucutre of the atmosphere, which still varies a great
deal within the hot Jupiter class.  Equilibrium chemistry suggests that high altitude ($P < 0.1$ bar) Fe and
silicate clouds may be present on the day-side of hot-Jupiters. In general,
cloud formation tends to increase the amount of scattered light
and smooth out many of the spectral features.  However at these altitudes the
atmosphere is purely radiative and likely fairly quiescent which would allow condensate particles
to quickly rain down to deeper levels of the atmosphere allowing only a relatively thin
haze to remain at high altitudes.  

 Hot-Jupiter temperature profiles can also enter a
high temperature, low pressure domain in which the molecules TiO and VO
that have strong optical absorption bands do not condense deeper in the atmosphere.
This leads to very strong heating by incident radiation and the formation of
exceptionally hot stratospheres akin to, but much hotter than, the stratospheres
driven by near-infrared methane absorption in the solar system giant planets  ({\em Hubeny et al.}, 2003;
{\em Fortney et al.}, 2006).

\section{\textbf{OBSERVATIONS OF HOT JUPITERS}}

Since their first, surprising detection a decade ago ({\em Mayor and Queloz}, 1995) the hot Jupiters have received substantial attention, leading to the detection of the planets both during transit and eclipse.  The chapter by Charbonneau et al. fully explores this topic.  Here we focus on the theoretical interpretation of the direct detections.

\subsection {Transmission Spectra}

As extrasolar planets transit their parent star, a small fraction of the
stellar flux passes tangentially through the planet's limb and upper
atmosphere.  The absorbing properties of the planetary atmosphere (along a
slant geometric path through the planet's limb) are added to the transmitted
stellar absorption spectrum.  There have been many published synthetic
transmission spectra for hot-Jupiters -- sometimes presented as the
wavelength-dependent planet radius that would be observed during a transit
event.  Some of these models assume plane-parallel slab geometry ({\em Seager and
Sasselov}, 2000; {\em Hubbard et al.}, 2001; {\em Brown}, 2001) while others assume spherical
geometry ({\em Barman et al.}, 2001, 2002).  All of these models have adopted a single
one-dimensional thermal profile intended to represent an  average of
the planet's limb.

{\em Seager and Sasselov} (2000) predicted strong transmission absorption features due
to Na and K alkali lines.  {\em Hubbard et al.} (2001) extended the modeling of
transmission spectra to near-IR wavelengths. Their models showed that, similar
to the Na and K alkali lines, H$_2$O bands can also imprint strong absorption
feature onto the transmitted spectrum.  Hubbard et al. also emphasized that
while the emission spectrum of the planet may have molecular bands diminished
by a reduced temperature gradient, the transmission spectrum is unaltered by
such affects.  

Modeling the transmission spectrum includes many potential difficulties.
Transmission spectroscopy probes the low pressure layers of the atmosphere
where non-equilibrium conditions are most likely to occur.  Also, the limb
(or the terminator) is the transition zone between the night side and the
irradiated day side.  Consequently, the stellar radiation passes through a
region that could have a steep {\em horizontal} temperature gradient and a
correspondingly steep gradient in the chemical composition along tangent path
lengths (see {\em Iro et al.}, 2005 and {\em Barman et al.}, 2005).  Such complications would
be difficult to represent accurately using a single one-dimensional model.
{\em Fortney} (2005) has also pointed out that, due to the relatively long tangential
path lengths, trace condensates (negligible to the emission spectrum) may have
a column density significant enough to impact the predicted transmission
spectrum.

\subsection{Thermal Emission}

Impressive new datasets appeared in  2005 that placed new constraints on hot-Jupiter atmospheres and stress-tested existing  hot-Jupiter atmosphere models.  Those planets which transit their primary stars are eclipsed by them half an orbital period later, during which time
only starlight--not planetary thermal emission--is detectable. The resulting light curve
yields the ratio of planet to stellar flux. Observations with {\it Spitzer Space Telescope} have constrained this 
ratio for both HD 209458b (at $24\,\rm\mu m$; 
{\em Deming et al.} 2005) and for TrES-1 (at 4.5 and $8 \,\rm\mu m$; 
{\em Charbonneau et al.} 2005). For HD 209458b, the known stellar flux can by multiplied by the flux ratio at
the same wavelength to yield the planetary flux,  $55 \pm 10\,\rm\mu Jy$, which can be expressed equivalently as a
brightness temperature of $1130 \pm 150\,\rm K$.  The distance to, and 
hence flux of, the TrES-1 star is not known.  Hence only a brightness 
temperature can be quoted with some certainty.  The TrES-1 temperature 
at $4.5\,\rm\mu m$ is $1010 \pm 60\,\rm K$ and at $8\,\rm\mu m$  is $1230 \pm 60\,\rm K$.
 
Despite only three thermal emission data points for two different 
planets, four model interpretations ({\em Barman et al.}, 2005; {\em Burrows et al.}, 2005; {\em Fortney et al.}, 2005; {\em Seager 
et al.}, 2005) have already been published!  Some of the published models have conflicting 
interpretations.  For example, {\em Burrows et al.} (2005) claim that the 
model interpretation of TrES-1 suggests that the planet is presently reradiating 
on the dayside, while the best fit models of {\em Fortney et al.} are those 
for which the incident stellar radiation is evenly redistributed.

Given such conflicting conclusions one might ask if the current data set is adequate to say anything 
concrete about the planetary atmospheres? Below we briefly summarize the currently published 
interpretations, and then provide our perspective on this question.  The intense interest in HD 209458b, however, does permit comparisons between groups modeling the same object with different approaches.  The range of a subset of the published models (Figure 4) provides a measure of the uncertainty at the current state of of the art.

{\em Burrows et al.} (2005) find that their predictions for the planet-to-star 
flux density ratios of both planets to be robust given the uncertainties in the planets' and
primary stars' physical properties.
 They inferred the presence of CO and perhaps $\rm H_2O$, and have 
determined that the atmospheres are hot. They suggest that the 
difference between the theoretical models and all three new 
measurements may be explained by an infrared-brighter hot dayside.

\begin{figure*}
 \epsscale{1}
 \plotone{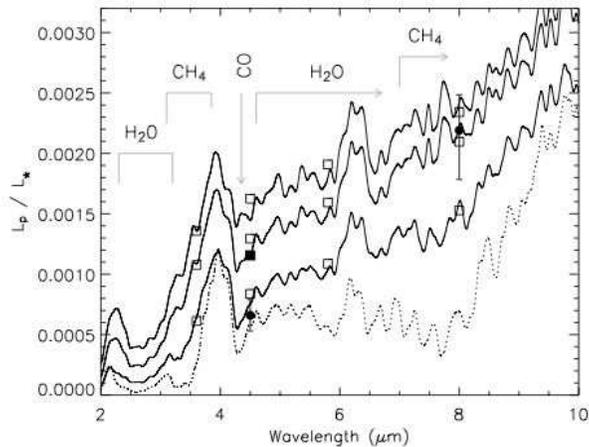}
  \caption{\small The ratio of the model flux from the day side of TrES-1 to that of its parent star, from {\em Barman et al.} (2005), assuming no
redistribution or $f=1$ (top curve) and redistribution models with $f = 0.5$
(middle solid curve) and $f = 0.25$ (bottom solid curve).  IRAC band fluxes for each model (found
by convolving with the IRAC response curves) are indicated with open squares
and filled circles show the {\it Spitzer} data with 1$\sigma$ error bars.  The
4.5 \micron\ IRAC value for a 10$\times$ solar, $f = 0.5$, model is also
shown (solid square).  The lower dotted line corresponds to an isolated brown
dwarf model with $T_{\rm eff} = 1150\,\rm K$. Note the much redder slope of the planet.  Major absorption bands are indicated.   }  
 \end{figure*}

{\em Fortney et al.} (2005) find that while standard solar metallicity models can fit the single datapoint
for HD 209458b, they do not for TrES-1, as the planetary spectral slope implied by the 4.5- and $8$-$\rm \mu m$ observations is redder than expected.  Model atmospheres that include a 3 to 5 times metal enhancement, or energy deposition into the atmosphere from 1-10 mbar, lead to a redder spectral slope.  With these models they find they can match the TrES-1 observations at $4.5\,\rm \mu m$ to 1 sigma, and at $8\,\rm \mu m$ to 2 sigma.  {\em Fortney et al.} find that the best-fit models for both planets assume that reradiation of absorbed stellar flux occurs over the entire planet.  They also note the to-date excellent agreement between {\it Spitzer} ultracool dwarf infrared spectral data and models ({\em Roellig et al.}, 2004).

In addition to standard solar abundance $f = 1/4$ and 1/2 models,
{\em Barman et al.}  (2005) compute the 2-D vertical {\em and}
horizontal temperature gradient over the entire day side in the
static no-redistribution ($f=1$) case.  For TrES-1, they find that
all three models are consistent with the  8-$\rm \mu m$ observations
at the 2 sigma level.  However, only their $f=1/4$ model agrees
with the 4.5-$\rm \mu m$ observation.  More importantly, agreement
between the $f=1$ model and the 4.5-$\rm \mu m$ observation would require
an unrealistic change to the atmospheric abundances.
Consequently, some process must be redistributing the absorbed
stellar flux at a level more comparable to an $f=1/2$ or 1/4
scenario.  This is in agreement with the findings of {\em Barman et
al.} and {Fortney et al.} (2005)  for HD209458b where an$ f = 1/4$ model
shows the best agreement with the 24 $\rm \mu m$ MIPS observations.
Figure 7 also illustrates just how red TrES-1 is compared to an
typical brown dwarf spectrum of the same luminosity.

Finally {\em Seager et al.} (2005) also conclude that a wide range of models fit the 
observational error bars. Starting with this philosophy, and including 
a $2.2$-$\rm \mu m$ observational upper limit ({\em Richardson et al.}, 2003) neglected by the other groups,  
they rule out some models for HD 209458b at 
the hot and cold end of the plausible temperature range.  
They show that models with $\rm C/O > 1$ can fit the HD 209458b data, 
including a paucity of $\rm H_2O$ (Figure 8) and describe how the same models 
could fit TrES-1.  They suggest that the models show an 
atmospheric circulation regime intermediate between pure {\it in situ} 
reradiation and very efficient heat redistribution.

\begin{figure*}
 \epsscale{1}
 \plotone{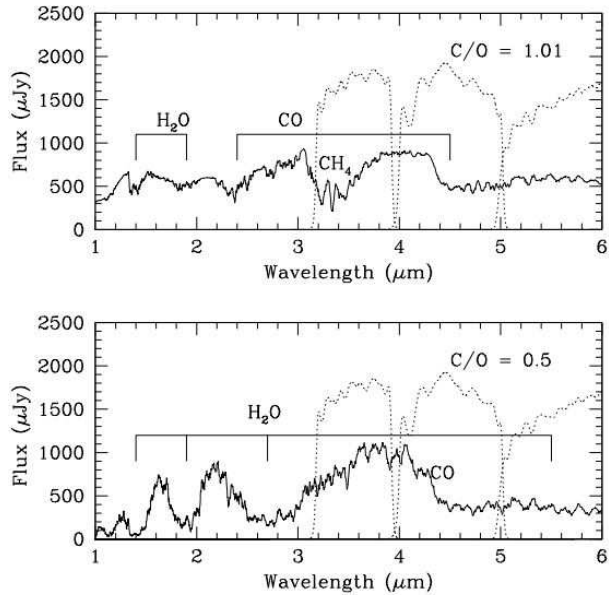}
  \caption{\small Thermal emission spectrum for HD209458b with C/O = 1.01 (and other 
elements in solar abundance). For C/O $>$ 1 at this planet's 
temperature, the water abundance is low and CH$_4$ abundance is 
increased, compared to a solar abundance spectrum (e.g., Figure 6). 
 For C/O greater than solar (0.5) but less than 1, the 
spectrum would still show reduced H$_2$O and increased CH$_4$ but to a 
lesser extent. The {\it Spitzer} IRAC bandpasses are shown as dotted lines.}  
 \end{figure*}

All modelers agree on one point: hot Jupiters are indeed hot. (Because the 
brightness temperatures of both HD 209458b and TrES-1 are over 1000 K, 
this conclusion does not, in fact, require a model interpretation at 
all.) The second point that all four modelers agree upon is that the 
TrES-1 4.5- and 8-$\rm \mu m$ data are not fit by a basic model with solar 
abundances: the model flux is too high in the 8-$\rm \mu m$ band compared 
to observations.  These two points are probably the only concrete 
inferences that can be made from the observations. The detailed 
arguments for validity of specific atmosphere conditions 
(stratosphere, clouds, photochemistry, etc.) must await further {\it Spitzer} data at other 
wavelengths.

\subsection{Non-equilibrium Effects}

So far nearly all hot-Jupiter models have assumed local thermodynamic
equilibrium (LTE) when solving the radiative transfer equation.  Assuming LTE
is tantamount to assuming that all species have level populations given by the
Saha-Boltzmann distribution and that the frequency- and depth-dependent source
function is simply a black body.  Consequently, potentially important effects
like photo-ionization and non-Boltzmann-like level populations are ignored.
Given that a large fraction of the radiation field in a hot-Jupiter atmosphere
is a non-local phenomena, LTE may be a rather risky assumption.  {\em Yelle} (2004) reviews
these and other issues related to understanding the likely very hot upper atmospheres of the
hot Jupiters.

Shortly after the detection of Na D absorption in the atmosphere of
HD209458b ({\em Charbonneau et al.}, 2002), {\em Barman et al.} (2002) and {\em Fortney et al.} (2003)
explored the possibility that non-LTE effects could alter the predicted
strength of Na absorption in hot-Jupiter atmospheres.  One possibility is that
Na is ionized to pressures less than about $0.01\,\rm bar$.  Also, the relevant level populations of
Na may be underpopulated, thereby reducing Na D line strengths.  

\subsection{Horizontal Gradients}

Unlike the atmospheres of isolated brown dwarfs, which are heated entirely from
the inside out, hot Jupiters experience significant heating by both internal
(heat leftover from formation) and external (heat from the star) sources of
energy.  The presence of this external source of energy breaks the spherical
symmetry implicitly assumed when one-dimensional model atmospheres are compared
to observations of brown dwarfs and stars, for example.  A lack of symmetry and
dual heating sources bring a number of challenging issues to the forefront of
the hot-Jupiter model atmosphere problem.  Indeed the day-night asymmetry and the need for more sophisticated modeling was
recognized very early on, as discussed in Section 2.1.

{\em Barman et al.} (2005) have made approximate 2-D static radiative-convective
equilibrium models for the day-side atmosphere for HD209458b and TrES-1.  These
models estimate the horizontal temperature gradients--in the absence of winds--to 
be quite steep ($\sim 1000$ K at $P = 1$ bar) and can lead to a complex
chemistry gradient over the day side.  For example, near the terminator, CO can
potentially be replaced as the dominate carbon bearing molecule by CH$_4$.  Na
condensation may also become important in this region (see also {\em Iro et al.},
2005).

As they are likely unstable, the steep horizontal gradients in the {\em Barman et al.} (2005) model strengthen the
case for modeling the effects of global circulation.  How significant the
impact of such circulations would be on the atmospheric structure (and thus on
the emergent spectrum) depends largely on the depth at which the stellar flux
is absorb and the radiative and advective time-scales in this region ({\em Seager et al.},
2005).
 
The strong day-side irradiation has motivated several groups to model the
global atmospheric circulation currents in hot-Jupiter atmospheres.
Three-dimensional simulations for HD209458b suggest the possibility of strong
zonal winds approaching or surpassing the sound speed ($\sim 1$ km s$^{-1}$ winds) and a
significant displacement of the atmospheric hot spot ({\em Showman and Guillot},
2000; {\em Cooper and Showman}, 2005).  Additionally, Showman and Cooper (2005) demonstrated the impact global winds
could potentially have on the depth-dependent temperature structure.  Their
simulations predict low pressure temperature inversions not seen in static 1-D
atmosphere models. {\em Cho et al.} (2003), assuming a characteristic
wind speed of 400 m s$^{-1}$ for HD209458b, found localized jets and vortices
producing hot and cold regions differing by as much as 300K.  These localized
``hot spots'' in the {\em Cho et al.} simulations also move about the poles with
$\sim 25$-day periods.  {\em Menou et al.} (2003) applied the results of {\em Cho et
al.} (2003) to other short-period EGPs and concluded that these kinds of
circulation patterns are likely to be common among hot Jupiters.  Despite
predicting very different atmospheric flows, all of these simulations agree
that circulation currents will most likely reposition the atmospheric hot spot(s)
away from the substellar point.  Consequently, maximum and minimum IR fluxes
would not necessarily coincide with orbital phases that align the substellar
and antistellar points with Earth, a result that can be tested by infrared light curves or, more easily, by secondary eclipse diagnostics ({\em Williams et al.}, 2006).

While the various hydrodynamic simulations differ significantly in details, there appears to be agreement that variations on the order of $\sim 300$ to 500 K may be present at ``photospheric'' pressures.  The studies of {\em Showman and Guillot} (2001) and more recently {\em Cooper and Showman} (2005) predict steady eastward supersonic winds producing an atmospheric hot spot that may be displaced $60^\circ$ from the planet's substellar point.


\section{\bf THE FUTURE}

Transiting planets and hot Jupiters are the most favorable extrasolar 
planets for observational study in the near future. They can be 
studied in the combined light of the planet-star system without direct 
imaging. Although ground-based observations will surely continue, the 
stable conditions of space make it the best place for the requisite 
high-precision observations. 
 
The transiting planets orbiting the brightest stars, particularly
HD209458b (the touch stone for hot Jupiters) as well as the
newly discoved HD189733b, will certainly continue to receive great attention.
{\em Spitzer} photometric and spectral observations will determine
if there are phase variations in the thermal emission and will search for spectral signatures of the atmosphere.
At visible wavelengths, the Canadian {\it MOST} 
space telescope will observe HD 209458b during secondary eclipse in 
scattered light and will reach a geometric albedo of 0.15. HST data 
will add to the variety of data on the same planet.  

Other non-transiting hot Jupiters will likely be  monitored with {\em Spitzer} 
for phase variation with both IRAC and MIPS. New transiting planets 
around bright stars, such as TrES-1 and HD149026 will also be observed 
with {\it Spitzer} and {\it HST}. {\it SOFIA} and large ground-based observatories such 
as Keck and VLT may be 
able to detect thermal emission at shorter wavelengths than Spitzer 
(2-$4\,\rm \mu m$) during secondary eclipse of transiting hot Jupiters. 
 
The more distant future for extrasolar planet characterization looks 
even more promising. In the next decade {\it Kepler} will find dozens of 
transiting extrasolar giant planets at a variety of semi-major axes 
out to 1 AU. Some of those planets' atmospheres may be detectable by 
{\em JWST}. Further to the future, in the later 
part of the next decade, planned 20- to 30-m ground-based 
telescopes may be able to directly image massive Jupiters in Jupiter-like orbits. 
Around the same time, NASA's {\it Terrestrial Planet Finders} and ESA's {\it Darwin} aim to directly detect
and characterize nearby extrasolar planets ranging from giant planets down to Earth-size planets.
  
As the era of extrasolar giant planet characterization moves from the hot Jupiters to the realm of true Jupiter analogs, the focus of characterization will change.  Mass and particularly radius, which are most easily constrained by transits, will be much less well constrained for more distant planets.  Instead spectra, which are especially sensitive to cloud structure, atmospheric composition, and atmospheric chemistry, will provide the primary method for planet characterization.  Ultimately obtaining the equivalent of the compositional fingerprint shown in Figure 1 for many planets in many planetary systems  will  illuminate the giant planet formation process as a function of stellar mass, metallicity, and planetary system architecture.

\bigskip

\textbf{ Acknowledgments.} This work was partially supported by the NASA Astrophysics Theory and Origins programs.  We thank  Heidi Hammel, Kevin Zahnle, and the anonymous referee for thoughtful suggestions that improved the manuscript as well as  Julie Moses for providing advice on photochemical issues. 

\bigskip

\centerline\textbf{REFERENCES}
\bigskip
\parskip=0pt
{\small
\baselineskip=11pt

\refs Ackerman A.~S. and Marley M.~S.\ (2001) {\em \apj, 556}, 872-884. 

\refs Allende Prieto C., Lambert, D.~L., and Asplund M.\ (2002) {\em \apj, 573,} L137-L140. 
 
\refs Appleby J.~F.\ (1986) {\em Icarus, 65}, 383-405. 

\refs Appleby J.~F. and Hogan J.~S.\ (1984) {\em Icarus, 59}, 336-366. 

\refs Asplund M.\ (2005) {\em Ann. Rev. Astron. Astrophys., 43}, 481-530.

\refs Atreya S.~K., Mahaffy, P.~R., Niemann H.~B., Won, M.~H., and Owen T.~C. (2003) {\em \planss, 51}, 105-112. 

\refs Baraffe I., Chabrier, G., Barman T.~S., Allard F., and Hauschildt P.~H.\ (2003) {\em \aap, 402}, 701-712.
  
\refs Barman T.~S., Hauschildt P.~H., and Allard F.\ (2001) {\em \apj, 556,} 885-895. 

\refs Barman T.~S., Hauschildt, P.~H., Schweitzer A., Stancil, P.~C., Baron E., and Allard F.\ (2002) {\em \apj, 569}, L51-L54.

\refs Barman T.~S., Hauschildt P.~H., and Allard F.\ (2005) {\em \apj, 632,} 1132-1139.

\refs Bodenheimer P., Lin D.~N.~C., and Mardling R.~A.\ (2001) {\em \apj, 548}, 466-472. 

\refs Bodenheimer P., Laughlin G., and Lin D.~N.~C.\ (2003) {\em \apj, 592}, 555-563.

\refs Brown T.~M.\ (2001) {\em \apj, 553}, 1006-1026. 

\refs Buriez J.~C. and de Bergh C.\ (1981) {\em \aap, 94}, 382-390.

\refs Burkert A., Lin D.~N.~C., Bodenheimer P.~H., Jones C.~A., and Yorke H.~W.\ (2005) {\em \apj, 618}, 512-523.

\refs Burrows A., et al.\ (1997) {\em \apj, 491}, 856-875.

\refs Burrows, A., Hubbard, W. B., Lunine, J. I., Marley, M. S., and Saumon, D. (2000) In {\em Protostars and Planets IV} (V.~Mannings, et al., eds). pp. 1339-1361, Univ. of Arizona, Tucson.

\refs Burrows A., Hubbard,  W.~B., Lunine, J.~I., and Liebert, J.\ (2001) {\em Reviews of Modern Physics, 73}, 719-765. 

\refs Burrows A., Sudarsky, D., and Hubbard, W.~B.\ (2003) {\em \apj, 594}, 545-551. 

\refs Burrows A.\ (2005) {\em \nat, 433,} 261-268. 

\refs Burrows A., Hubeny I., and Sudarsky D.\ (2005) {\em \apj, 625}, L135-L138. 

\refs Chabrier, G., Barman, T., Baraffe, I., Allard, F., and Hauschildt, P.~H.\ (2004) {\em \apj, 603}, L53-L56. 

\refs Charbonneau D., Brown T.~M., Noyes R.~W. and Gilliland R.~L.\ (2002) {\em \apj, 568}, 377-384.

\refs Charbonneau D., et al.\ (2005) {\em \apj, 626}, 523-529.
 
\refs Cho J.~Y.-K., Menou K., Hansen B., Seager S.\ (2003) {\em \apj, 587}, L117-L120.

\refs  Cooper C.~S., Sudarsky D., Milsom J.~A., Lunine J.~I., and Burrows A.\ (2003) {\em \apj, 586}, 1320-1337. 

\refs Cooper C.~S. and Showman A. (2005) {\em \apj, 629}, L45-L48.
 
\refs Dlugach J.~M. and Yanovitskij E.~G.\ (1974), {\em Icarus, 22}, 66-81. 
 
\refs Dyudina U.~A., Sackett P.~D., Bayliss D.~D.~R., Seager S., Porco C.~C., Throop H.~B., and Dones L.\ (2005) {\em \apj, 618}, 973-986.

\refs Fegley B. and Prinn R.~G.\ (1988) {\em \apj, 324}, 621-625.

\refs Fegley B.~J., Gautier D., Owen T., and Prinn R.~G.\ (1991) In {\it Uranus} (J.~T.~Bergstralh, E.~D.~Miner, and M.~S.~Matthews, eds.), pp. 147-203. Univ.~of Arizona, Tucson.
 
\refs Fegley B.~J. and Lodders K.\ (1994) {\em Icarus, 110}, 117-154.
 
\refs Flasar F.~M., et al.\ (2005) {\em Science, 307}, 1247-1251. 

\refs Fortney J.~J.\ (2005)  {\em \mnras, 364}, 649-653.
 
\refs Fortney J.~J., Marley M.~S., Lodders K., Saumon D., and Freedman R.\ (2005) {\em \apj, 627}, L69-L72. 

\refs Goukenleuque C., B{\'e}zard B., Joguet B., Lellouch E., and Freedman R.\ (2000) {\em Icarus, 143}, 308-323. 

\refs Griffith C.~A. and Yelle R.~V.\ (1999) {\em \apj, 519}, L85-L88.
 
\refs Guillot T., Burrows A., Hubbard W.~B., Lunine J.~I., and Saumon D.\ (1996) {\em \apj, 459}, L35-L38.

\refs Guillot T. and Showman A.~P.\ (2002) {\em \aap, 385}, 156-165.

\refs Helling C., Klein R., Woitke P., Nowak U., and Sedlmayr E.\ (2004) {\em \aap, 423}, 657-675.
 
\refs Hogan J.~S., Rasool S. I. and Encrenaz, T. (1969) {\em J. Atmos. Sci., 26}, 898-905.

\refs Hubbard W.~B.\ (1968) {\em \apj, 152}, 745-754.

\refs Hubbard W.~B.\ (1977) {\em Icarus, 30}, 305-310.

\refs Hubbard W.~B., Fortney J.~J., Lunine J.~I., Burrows A., Sudarsky D., and Pinto P.\ 
(2001) {\em \apj, 560}, 413-419.

\refs Hubeny I., Burrows A., and Sudarsky D.\ (2003) {\em \apj, 594}, 1011-1018. 
 
\refs Iro N., B{\'e}zard B., and Guillot T.\ (2005) {\em \aap, 436}, 719-727. 

\refs Kuiper G.~P.\ (1952) {\em The Atmospheres of the Earth and Planets.} University of Chicago Press, Chicago.

\refs Kurucz R. (1993) CD-ROM 13, ATLAS9 Stellar Atmosphere Programs and
2 km/s Grid (Cambridge: SAO).

\refs Laughlin G., Wolf A., Vanmunster T., Bodenheimer P., Fischer D., Marcy G., Butler P., and 
Vogt S.\ (2005) {\em \apj, 621}, 1072-1078. 

\refs Liang M.-C., Seager S., Parkinson C.~D., Lee A.~Y.-T., and Yung Y.~L.\ (2004) {\em \apj, 605}, L61-L64. 

\refs Lissauer J.~J.\ (1993) {\em Ann. Rev. Astron. Astrophys., 31}, 129-174.

\refs Lodders K.\ (2003) {\em \apj, 591}, 1220-1247.

\refs Lodders K.\ (2004) {\em \apj, 611}, 587-597. 

\refs  Marley M.~S., Saumon D., Guillot T., Freedman R.~S., Hubbard W.~B., Burrows A., and Lunine 
J.~I.\ (1996) {\em Science, 272}, 1919-1921.

\refs Marley M.~S.\ (1998) In {\it Brown Dwarfs and Extrasolar Planets} (R.~Rebolo et al., eds.), pp. 383-393. ASP Conf. Series, San Francisco.

\refs Marley M.~S. and McKay C.~P.\ (1999) {\em Icarus, 138}, 268-286.

\refs Marley M.~S., Gelino C., Stephens D., Lunine J.~I., and Freedman R.\ (1999) {\em \apj, 513}, 879-893. 

\refs Marley M.~S., Seager S., Saumon D., Lodders K., Ackerman A.~S., Freedman R.~S., and Fan X.\ (2002) {\em \apj, 568}, 335-342. 

\refs Mayor M. and Queloz D. (1995) A Jupiter-mass companion to a solar-type 
star. {\em Nature, 378}, 355-359. 

\refs Menou K., Cho J.~Y.-K., Seager S., Hansen B.\ (2003) {\em \apj, 587}, L113-L116.

\refs Owen T., Mahaffy P., Niemann H.~B., Atreya S., Donahue T., Bar-Nun A., and de Pater I.\ (1999) {\em \nat, 402}, 269-270.

\refs Pollack J. B., Hubickyj O., Bodenheimer P., Lissauer J. J., Podolak 
M., and Greenzweig Y. (1996)  {\em Icarus, 124}, 62-85. 

\refs Prinn R.~G. and Barshay S.~S.\ (1977) {\em Science, 198}, 1031-1034. 

\refs Richardson L.~J., Deming D., Wiedemann G., Goukenleuque C., Steyert D., Harrington J., 
and Esposito L.~W.\ (2003) {\em \apj, 584}, 1053-1062. 

\refs Roellig T.~L., et al.\ (2004) {\em \apjs, 154}, 418-421.

\refs Rossow W.~B.\ (1978) {\em Icarus, 36}, 1-50.
 
\refs Saumon D., Marley M., and Lodders K. (2003) {\em arXiv:astro-ph/0310805}.

\refs Seager S. and Sasselov D.~D.\ (1998) {\em \apj, 502}, L157-L161.

\refs Seager S., Whitney B.~A,. and Sasselov D.~D.\ (2000) {\em \apj, 540}, 504-520. 

\refs Seager S., Richardson L.~J., Hansen B.~M.~S., Menou K., Cho J.~Y.-K. and Deming D.\ (2005) {\em \apj, 632}, 1122-1131.

\refs Showman, A. and Cooper, C.~S. (2006) In {\it Tenth Anniversary of 51 Peg-b:
Proc. of Haute Provence Observatory Colloq.} (L.~Arnold, F.~Bouchy, and C.~Moutou, eds.), in press.

\refs Sudarsky D., Burrows  A., and Pinto P.\ (2000) {\em \apj, 538}, 885-903. 

\refs Sudarsky D., Burrows A., and Hubeny I.\ (2003) {\em \apj, 588}, 1121-1148. 

\refs Sudarsky D., Burrows A., Hubeny I., and Li A.\ (2005) {\em \apj, 627}, 520-533. 

\refs  Trafton L.~M.\ (1967) {\em \apj, 147}, 765-781. 

\refs Tsuji T.\ (2005) {\em \apj, 621}, 1033-1048. 

\refs Vidal-Madjar  A., Lecavelier des Etangs A., D{\'e}sert J.-M., Ballester G.~E., Ferlet R., 
H{\'e}brard G., and Mayor M.\ (2003) {\em \nat, 422}, 143-146.
 
\refs Visscher C. and Fegley B.~J.\ (2005) {\em \apj, 623}, 1221-1227. 

\refs Westphal J.~A., Matthews K., and Terrile R.~J.\ (1974) {\em \apj, 188}, L111-L112.

\refs Wetherill G.~W. and Stewart G.~R.\ (1989) {\em Icarus, 77}, 330-357.
 
\refs Yelle R. (2004) {\em Icarus, 170}, 167-179.

\refs Young R.~E. (2003) {\em New Astron. Rev., 47}, 1-51.

\refs Yung Y.~L., Drew W.~A., Pinto J.~P. and Friedl R.~R.\ (1988) {\em Icarus, 73}, 516-526.

\end{document}